\input{aipcheck}

\documentclass[
    ,final            
  ]
  {aipproc}

\layoutstyle{6x9}

\begin{document}

\title{A Simple Method To Test For Energy-Dependent Dispersion In High Energy Light Curves Of Astrophysical Sources}

\author{M.~K.~Daniel}{
  address={Department of Physics, University of Durham, Durham, DH1 3LE. U.K.}
}

\author{U. Barres de Almeida}{
  address={Max-Planck-Institut f\"{u}r Physik, D-80805, M\"{u}nchen, Deutschland.}
}

\begin{abstract}
We present a method of testing for the presence of energy dependent 
dispersion in transient features of a light curve. It is based on 
minimising the Kolmogorov distance between two cumulative event 
distribution functions. The unbinned and non-parametric nature of the 
test makes it particularly suitable for searches of statistically 
limited data sets and we also show that it performs well in the presence 
of modest energy resolutions typical of gamma-ray observations 
($\sim20\%$). We illustrate its potential to set constraints on 
quantum-gravity induced Lorentz invariance violation effects from 
observations by the current and future generation of ground-based 
gamma-ray telescopes.
\end{abstract}

\maketitle


\section{Introduction}
Timing analysis algorithms with the capability of resolving energy dependent
properties can be an important tool for probing the physical mechanisms leading
to flux variability, such as particle acceleration and cooling \cite{Bednarek08}, 
or the nature of a propagating medium \cite{Amelino-Camelia98}. In the case of
very high energy gamma-ray sources, where high energy processes can
be responsible for extreme and short-lived variability events, the observational
data are often limited by low photon statistics and non-negligible uncertainty
in the reconstructed energy of indivdual events. This makes unbinned methods,
which act on the information of the entire available sample, the natural and
preferential choice of approach to the temporal analysis of these event lists.

\section{Method}
\begin{figure}[hp]
  \includegraphics[width=\textwidth]{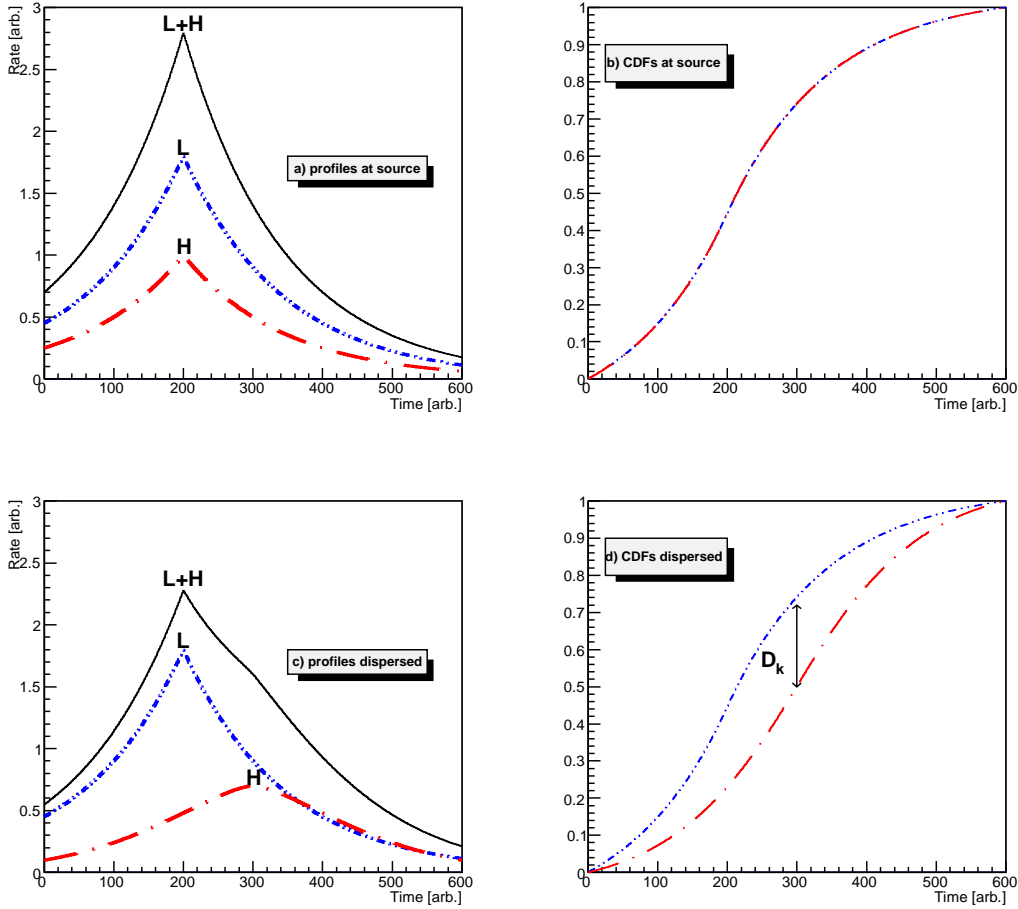} 
  \caption{Cartoon of the effect of the energy dependent dispersion on 
           the shape of the low (L) and high (H) energy profiles. The 
           panels on the left show the shape of the lightcurve and the 
           panels on the right the event CDFs. The top plots show are 
           the intrinsic (at source) shape and the bottom after 
           propagation.}
  \label{fig:cartoon}
\end{figure}

If the low ($L$) and high ($H$) energy particles are generated in the 
same region then they must be able to exist co-spatially. The act of 
acceleration, or cooling, or moving through a dispersive medium will act 
to separate the $L$ and $H$ populations relative to each other. An 
energy dependent correction factor ($\tau$) can be applied to the event 
arrival times ($t_{i}$)
\begin{equation}
 \delta t_{i} = - \tau E_{i}^{\alpha}
\end{equation}
where $\delta t$ is the difference in arrival time with and without 
dispersion, $E_{i}$ is the energy of the event and $\alpha$ is the 
scale of the correction (1 for linear, 2 for quadratic, etc). By cycling 
through a range of correction factors we can determine the one 
($\tau^{*}$) where the shape of the $H$ light-curve best fits that of 
the $L$ one, here we use the Kolmogorov distance between the cumulative 
distribution function (CDF) of the event arrival times, as seen in 
figure~\ref{fig:cartoon}.

\begin{figure}[hbp]
  \includegraphics[width=.45\textwidth]{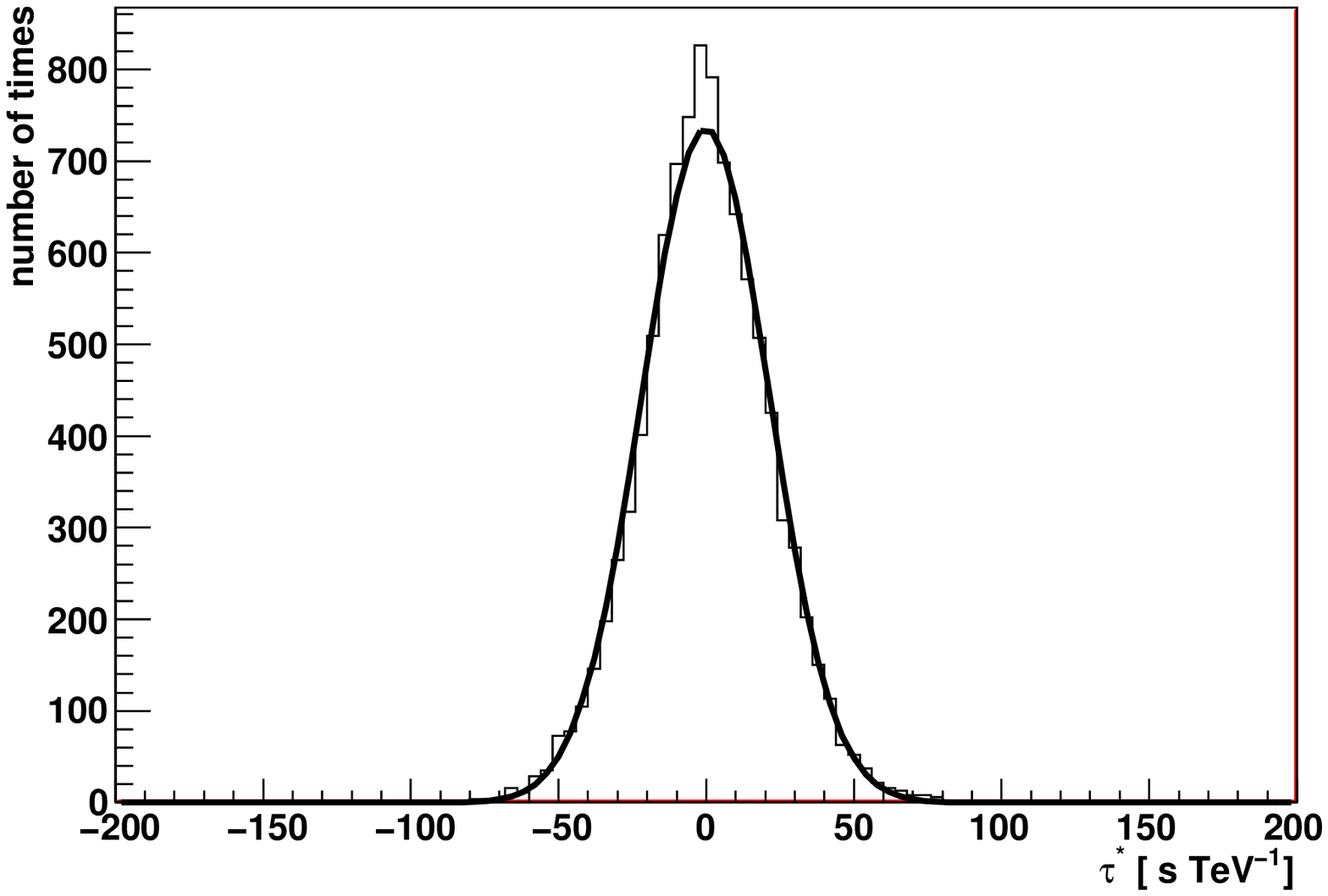}
  \includegraphics[width=.55\textwidth]{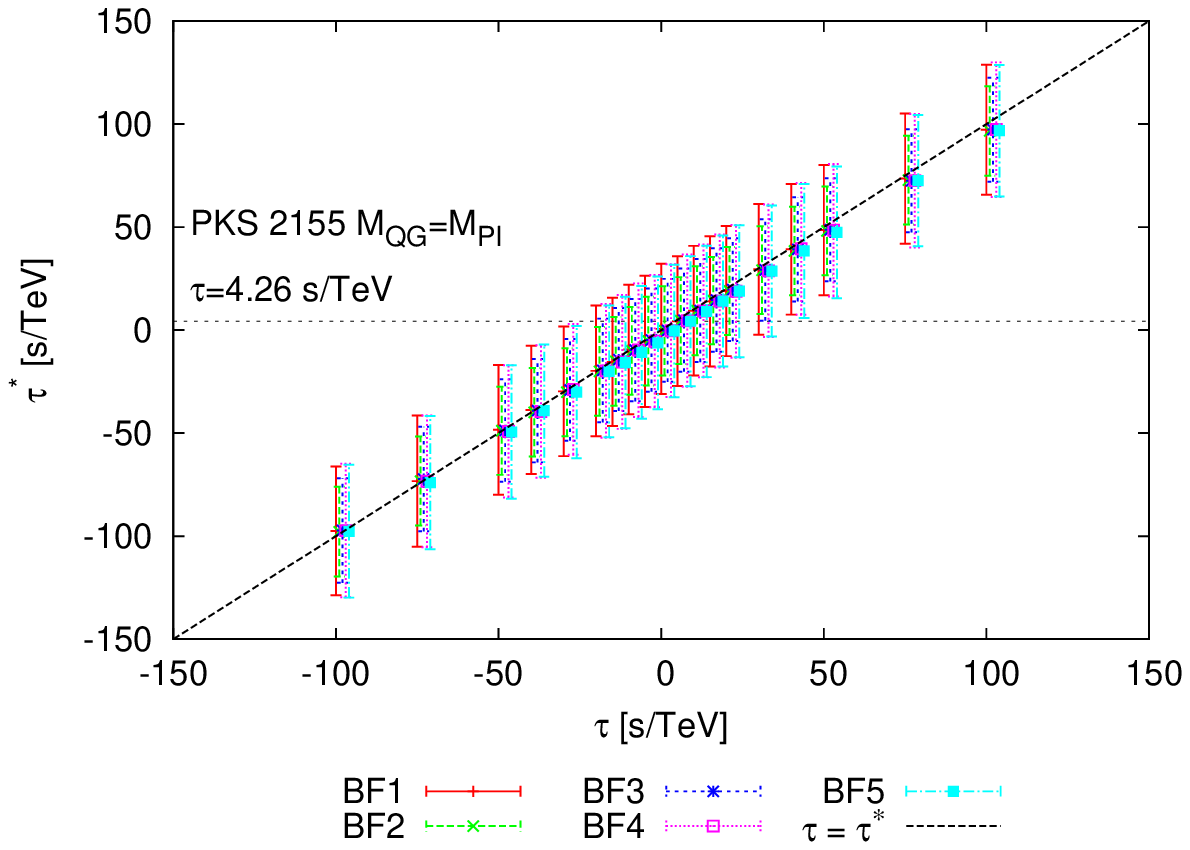}
  \caption{Performance of the method for recovering dispersion. 
           The left plot shows the error in the best estimate is well fit by a
           Gaussian; 
           the right plot shows the accuracy to which the estimated dispersion
           matches the actual simulated dispersion (see text for details).}
  \label{fig:perf}
\end{figure}

Simulating 10,000 lightcurves shows the Kolmogorov test always has a 
well defined minimum, with the difference between the expected and best 
correction ($\tau-\tau^{*}$) well fit by a Gaussian~\ref{fig:perf}. The RMS of 
the fit is dependent on the width of the light-curve, but relatively 
insensitive to the rise and fall times or the number of events contained within, 
provided there are $\geq 10$ events in the $H$ sample. It is also 
relatively insensitive to the energy binning provided the 
$\overline{E_{H}} \geq 2\overline{E_{L}}$.

\begin{figure}[htbp]
  \includegraphics[height=.45\textheight]{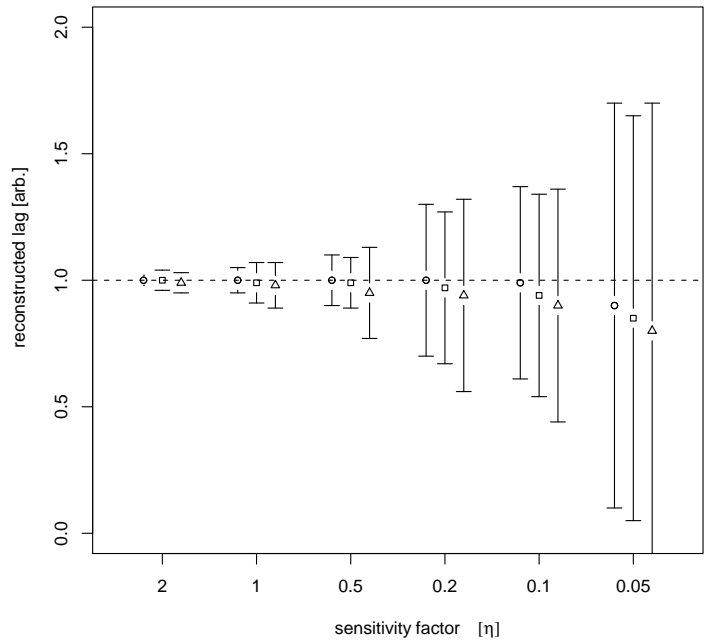}
  \caption{Sensitivity $\eta$ of the algorithm for $0\%$ (open circle), 
           $10\%$ (open square) and $20\%$ (open triangle) energy resolution.}
  \label{fig:Eres}
\end{figure}

We quantify the sensitivity to the burst width by the term sensitivity 
factor, $\eta$ defined as 
\begin{equation}
 \eta = \frac{\delta t}{\Delta t}
\end{equation}
where $\Delta t$ is the width of the transient feature in the 
light-curve. In figure~\ref{fig:Eres} we simulated 10,000 Gaussian burst 
profiles of 500 events each and a power-law spectral index of -2.5. A 
dispersion was introduced that varied from 5-200\% of the burst width. 
We see, as expected, that the narrower a burst relative to the 
dispersion the better it can be determined. Also plotted in 
figure~\ref{fig:Eres} are the results of varying energy resolutions 
($|\Delta E/E|$) from ideal (0\%, 10\% and 20\%). There is a small 
systematic trend for the reconstructed lag to be underestimated as the 
energy resolution worsens, again this is to be expected, but this is 
very small in comparison to the overall statistical error in $\tau^{*}$ 
showing the method is robust to the modest energy resolutions expected 
in ground based gamma-ray astronomy. It is possible to overcome this 
systematic trend with appropriate Monte Carlo modelling or 
bootstrapping, if necessary.

\section{Discussion \& Conclusions}
We have presented a simple method to perform an unbinned, non-parametric energy
dependent timing analysis of data with low statistics and moderate energy
resolutions. Further details of the performance of the method can be found in
\cite{BdA12}, simulations of current generation VHE gamma-ray instrument
observations of AGN show the method to be comparable in sensitivity to the more
sophisticated analyses which have to make greater assumptions on the instrinsic
source physics and instrument response functions. The placing of Planck scale
limits on the linear term in Lorentz invariance violation due to quantum gravity
models could be achievable in observations by the next generation instrument CTA 
\cite{Doro12}.





\bibliographystyle{aipproc}   




\end{document}